\PassOptionsToPackage{table}{xcolor}
\documentclass[sigconf]{acmart}
\AtBeginDocument{%
  }

\setcopyright{acmlicensed}
\copyrightyear{2025}
\acmYear{2025}
\acmDOI{3735079.3735321}
\acmConference[tbd. '25]{tbd.}{tbd.}{tbd/}
\acmISBN{978-1-4503-XXXX-X/2018/06}

\usepackage{listings}

\usepackage{xcolor}          

\usepackage{tabularx}
\usepackage{booktabs}

\begin{document}

\title{Metadata Management for AI-Augmented Data Workflows}

\author{Jinjin Zhao}
\email{j2zhao@uchicago.edu}
\affiliation{%
  \institution{University of Chicago}
  \city{Chicago}
  \state{IL}
  \country{USA}
}

\author{Sanjay Krishnan}
\email{skr@uchicago.edu}
\affiliation{%
  \institution{University of Chicago}
  \city{Chicago}
  \state{IL}
  \country{USA}
}


\begin{abstract}
AI-augmented data workflows introduce complex governance challenges, as both human and model-driven processes generate, transform, and consume data artifacts. These workflows blend heterogeneous tools, dynamic execution patterns, and opaque model decisions, making comprehensive metadata capture difficult. In this work, we present TableVault, a metadata governance framework designed for human-AI collaborative data creation. TableVault records ingestion events, traces operation status, links execution parameters to their data origins, and exposes a standardized metadata layer. By combining database-inspired guarantees with AI-oriented design, such as declarative operation builders and lineage-aware references, TableVault supports transparency and reproducibility across mixed human-model pipelines. Through a document classification case study, we demonstrate how TableVault preserves detailed lineage and operational context, enabling robust metadata management, even in partially observable execution environments.

\end{abstract}




\maketitle

\section{Introduction}


The complexity of data generation has grown dramatically as the number of frameworks, systems, and people involved in generating a single data artifact has expanded. In classical enterprise data, a single author entity oversaw the entire lifecycle of the dataset, designing the schema, establishing the collection protocols, enforcing correctness standards and often running the complete analysis pipeline. As the internet matured, semi-centralized dataset production emerged as organizations began to leverage external resources to expand the scale, richness, and timeliness of their data. Web scraping, social media monitoring, and crowdsourcing platforms like Amazon Mechanical Turk introduced new participants into the data production process.\cite{Franklin2011CrowdDBAQ}. While this dramatically increased the quantity and diversity of available data, it also introduced new risks: uneven data quality, hidden biases, and opaque collection contexts. Not surprisingly, web- and crowd- augmented dataset inspired seminal research on data provenance~\cite{Cheney2009ProvenanceID}, truth-finding~\cite{dong2015data}, and metadata management~\cite{Hellerstein2017GroundAD}.

We are now witnessing the rise of a new methodology, AI-augmented dataset production, where artificial intelligence tools are used to actively shape analytic data~\cite{shankar2025docetlagenticqueryrewriting,Zeng2023ScientificOS, Levine2024Cell2SentenceTL, Hosseini2023FightingRF, Abramson2024AccurateSP, Edge2024FromLT, Arora2023LanguageME, Fernandez2023HowLL, vanSchaik2024AFG, Anderson2024TheDO}. AI systems can clean and impute missing data, enrich raw inputs with labels or summaries, and even generate synthetic content to fill gaps in coverage. The AnnotatedTables project used a model to annotate 405,616 executable SQL queries \cite{hu2024annotatedtableslargetabulardataset}, and The 220k‑GPT4Vision Captions dataset attaches dense, automated natural‑language descriptions to about 218000 images from LVIS.\cite{wang2023believepromptinggpt4vbetter}.

While AI tools can semantically manipulate data at an unprecedented scale, we see a similar challenge in trust and integrity to those in web-augmented and crowd-augmented datasets. These challenges are further complicated, as human evaluation (through prompt adjustments and data reviews) is now intertwined with automated transformations. This raises questions about how to generate authenticity and accountability in this novel context. 

The FAIR principles, generated from prominent domain experts and data stakeholders, emphasize the importance of metadata in achieving good data practices. Specifically, it notes that metadata should be linked to the original data source, accessible to subsequent practitioners, contain detailed lineage information, and have a standardized format \cite{WilkinsonFAIR2016}. In the context of AI collaboration, we now must consider how to achieve these principles, given that an AI model, alongside humans, may both generate the initial workflow and be the subsequent practitioner that ingests metadata information. 

In this paper, we argue that process can be made by supporting metadata at the systems level. For this purpose, we propose an initial framework, TableVault, to improve metadata management in AI-augmented data workflows. TableVault is designed around four properties that improve the capture of metadata: (1) guaranteed record of data ingestion, (2) robust operation status tracing, (3) lineage capture of execution parameters, and (4) metadata querying within model execution. It accomplishes this by merging traditional database principles with ideas from modern AI practice. 

\begin{figure*}[t]
  \centering
  \includegraphics[width=0.65\linewidth]{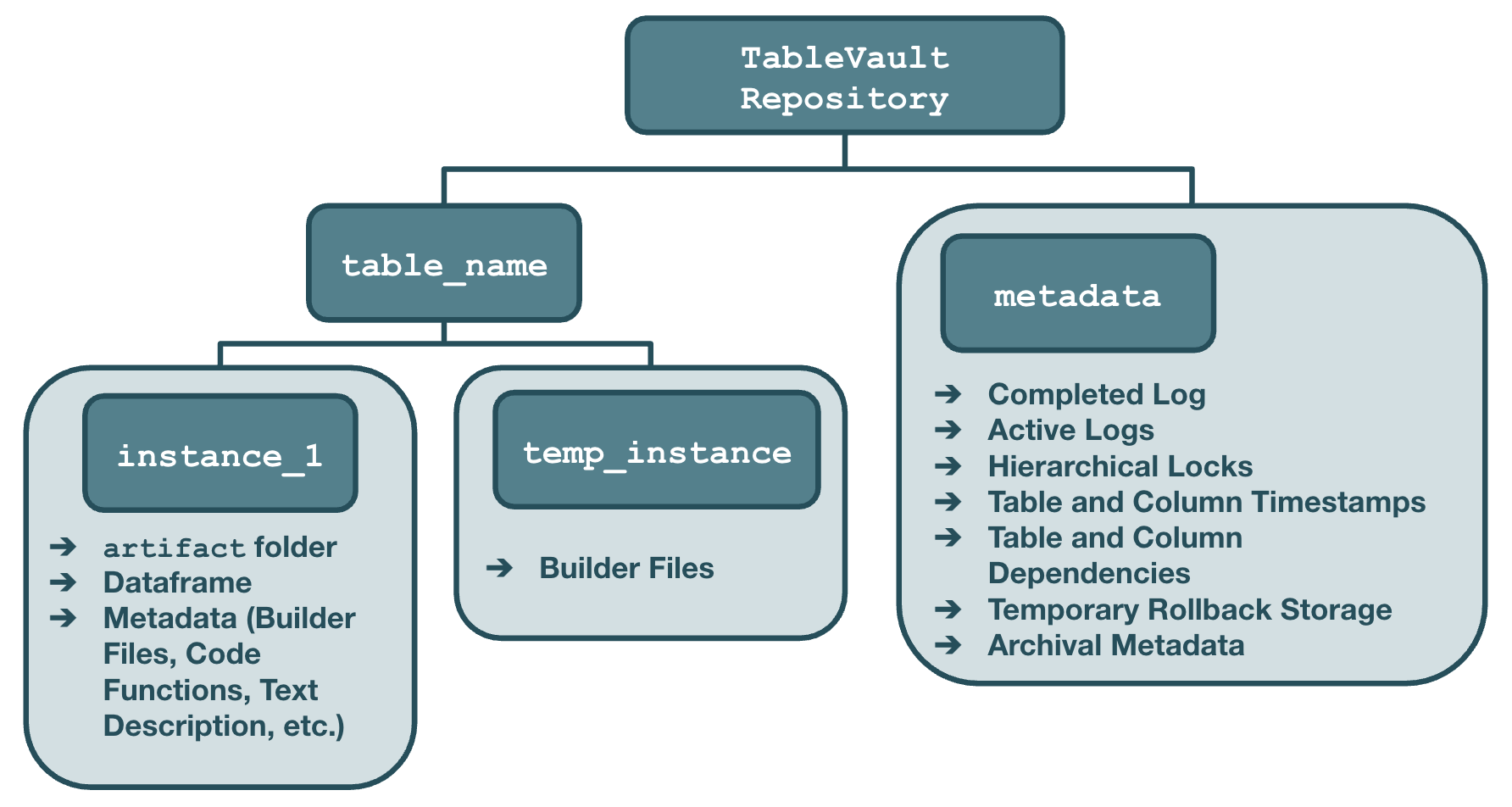}
  \caption{System diagram of TableVault.}
  \label{fig:system}
\end{figure*}

\textbf{Guaranteed Record of Data Ingestion} In contemporary data-processing workflows, it would be difficult to replicate the benefits of specialized tools with one system - for example, a generalized workflow system trying to beat a Jupyter notebook for quick adhoc data transformations \cite{Kluyver2016Jupyter}.  However, achieving comprehensive, end-to-end metadata capture across a heterogeneous toolchain is impracticable, as it would require deep instrumentation of every external component. Instead, we argue for a practical opportunistic approach, where full metadata is traced when operations occur within the system, and data ingestion points are recorded to signal when external operations may have occurred. The consequence of this approach is that it requires \emph{a single overarching system to have oversight of top-level workflow orchestration, as well as control over data storage and versioning}.
    
\textbf{Robust Operation Status Tracing.} Even for operations executed within the system, the capture of data lineage is complex. Language models make data transformations not only based on the input data, but also based on their internal knowledge of the world \cite{Petroni2019LanguageMA}. Additionally, the relationship between operation input and output may not be transparent. Therefore, we believe that full lineage may not be completely traced by the execution system but rather be captured by workflow-dependent explainability tools. To account for this change, a metadata system should \emph{support application-level explainability by linking its partially-known data lineage to exact operation executions}. 
        
 \textbf{Lineage Capture of Execution Parameters.} Execution plans and configuration parameters are increasingly  driven by data \cite{Marcus2021Bao, Snoek2012PracticalBO, Shin2020AutoPrompt}. For instance, processes like hyperparameter tuning, prompt engineering, and query optimization may be determined by pre-existing data or experiments, rather than a handcrafted algorithm. Although this optimization may occur with an execution system, we believe that it is natural to allow this customization within the workflows themselves. This shift demands that \emph{lineage capture extends beyond direct data artifacts to include the data inputs that influenced the generation of the execution logic itself.}

 \textbf{Metadata Querying Within Model Executions.}  The emergence of model-based agents capable of executing end-to-end tasks means that the data workflow itself may be dynamic \cite{yao2023reactsynergizingreasoningacting}. To enable this new type of workflow, we propose that the system should allow agents within a workflow to have direct access to the full metadata layer. This requires \emph{the metadata to be organized in a standardized structure and can be programmatically queried within operation executions}. The execution of these queries themselves may be recorded for future introspection.



\section{How TableVault Captures Metadata}

TableVault is responsible for both the operation execution and data artifact management in human-model data workflows; it has clear oversight over metadata of all artifacts generated throughout the entire data creation process. The system is implemented for the Python programming language and operates over structured dataframes and collections of arbitrary unstructured data artifacts. Dataframes are a popular structure and enable many common operations found in data science \cite{petersohn2020scalabledataframesystems}. 

Each data operation executed in TableVault is linked to an author, which may be an human operator or a parent operation process. The author has direct access to the operational state, metadata, and partial data of an in-process operation. However, subsequent operations that directly depend on the generated data only observes the final generated version of an instance, which is immutable and contains structured metadata about the instance's data lineage and operation status.

\subsection{System Storage Architecture}

Let us briefly describe how data is internally stored in a TableVault repository. In our current implementation, all operational information and data artifacts are stored in human-readable files on the native file system. Figure \ref{fig:system} shows the organizational structure of these files. At the outermost level, a TableVault repository contains a collection of ``tables''. In TableVault, a \texttt{table} is a semantic collection of instances that are stored together in the same folder. When retrieving data within TableVault, the \texttt{table\_name} string references to the latest instance in a table. 

A TableVault instance represents a dataframe, with its associated metadata and optional unstructured artifacts. For example, one instance may contain a collection of images inside its \texttt{artifact} folder along with an indexing dataframe that has name references to each image in the collection. In the same repository, another instance may hold a dataframe of the categorical names generated for each image reference. This data organization supports popular document processing workflows \cite{shankar2025docetlagenticqueryrewriting}. 

In terms of the exact metadata recorded, each instance contains the configuration YAML files (builders) and the executed Python code files that generated the table. All instances and enclosing tables also contain a descriptive YAML file with generated data lineage information and optional user descriptions. In the metadata folder, there exist separate logs for completed and currently active operations. Upon deletion of any table or instance, only the dataframe and artifacts are deleted, and the associated metadata is moved to an archival location \cite{WilkinsonFAIR2016}.

\subsection{Managed Data Operations} 

In a language model operation, the exact dependency between input parameters and output values is often opaque. For example, two different prompts may result in two different output datasets, but it may not be apparent which dataset matches to which prompt. In other words, external verification of execution may not be feasible, and the underlying execution system must explicitly record the operation execution status. This is not trivial since, to support data-driven execution planning, operations may be paused, reverted, or restarted dynamically. Hence, despite  being primarily a metadata system, TableVault has internally implemented an extensive framework for general data operation execution.

The creation and deletion of tables, instances, and dataframes is carefully tracked within the TableVault API. We establish ACID-like properties through standard database practices, including techniques such as two-phase locking and write-ahead logs \cite{GrayReuter1993}. To guarantee atomicity, the system creates hard links on potentially overwritten files at setup, and then uses a copy-on-write strategy during the operation execution. For consistency, operation-specific validations are leveraged, and the operation is automatically reverted if any checks are not passed. To achieve isolation while enabling multiple concurrent users, during the setup phase, operation processes can request shared or exclusive locks at the table and instance level. Finally, the system is inherently durable since all state data (including locks) are written to the file system.

On top of these techniques, TableVault gives the author direct access to the data and metadata files of partial operations, and allows user input for error-handing, interrupts, and restarts. When an operation is interrupted, it can either be completely reverted, or it can maintain its locks and persists in a paused state until the author resumes or stops the operation. Each decision is recorded in the system metadata, and lineage between the author and operation state is included in the record. 

For processes that are not direct authors of the operation, any data generated by the operation is not accessible until it is completed and its status is recorded. This ensures that the accessed data of an instance is consistent with its final recorded metadata.

\subsection{Data Creation in TableVault}

There are two specific paths to create a dataframe and associated artifacts in TableVault. An author may execute an Python operation within TableVault so that it is fully managed, or they may import externally formed data into the system. If the operation is executed within TableVault, it uses our declarative interface that enables explicit data input from existing TableVault instances for all input parameters.

\begin{figure}[htbp]
  \centering
  \includegraphics[width=\linewidth]{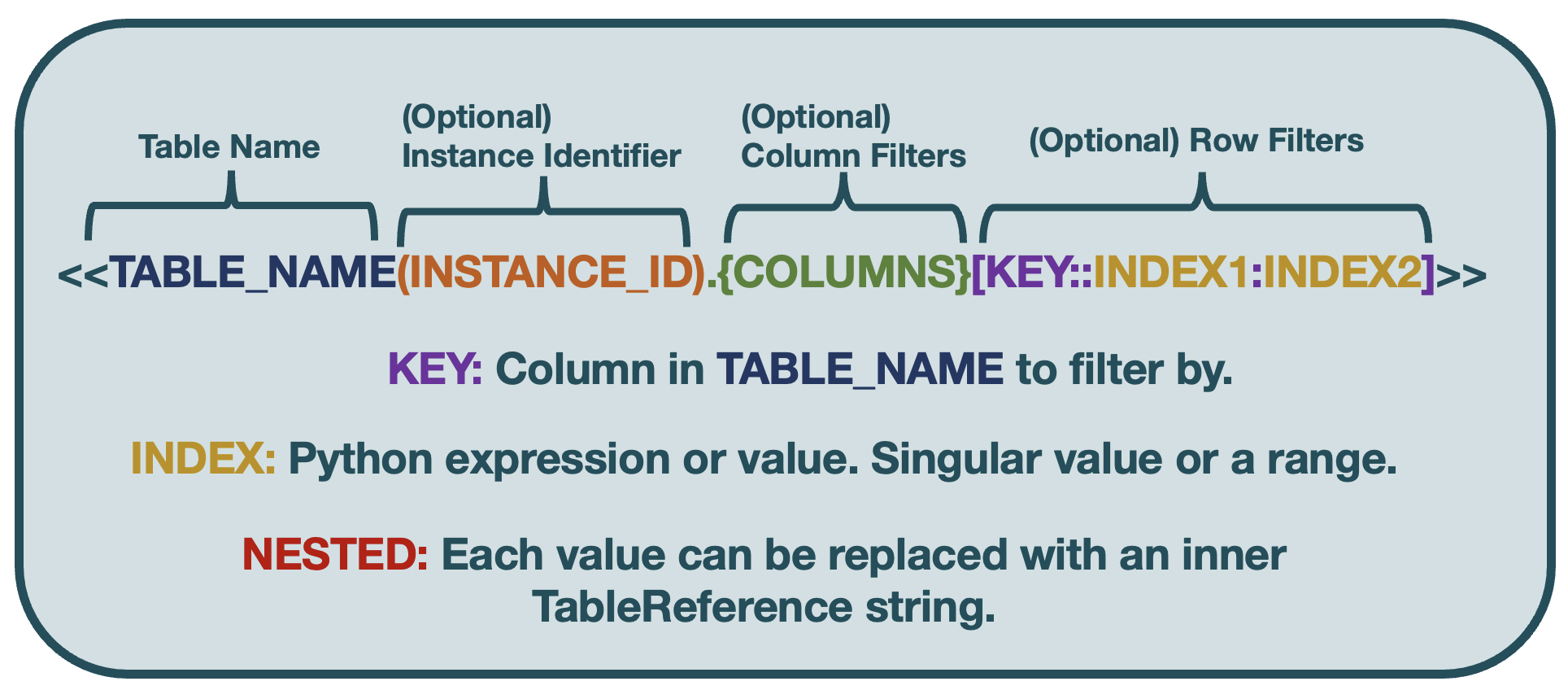}
  \caption{Breakdown of a TableReference string.}
  \label{fig:table_reference}
\end{figure}

It has become common practice to parameterize language model operations with a builder YAML file \cite{shankar2025docetlagenticqueryrewriting}. While this doesn't change system implementation, a key design decision in TableVault is that this file exposes execution parameters to the author, and unifies parametrization for execution and operations. This allows for complex data-driven workflows, is that each of these parameters can be linked to external instances in the same repository. 

We have implemented a custom string data type called a \texttt{Table- Reference} string to match TableVault's data structure. These strings are designed to follow Pythonic index conventions\cite{harris2020array} and can reference other full and partial dataframes. Figure \ref{fig:table_reference} shows the standard form of a \texttt{TableReference} string. A instance is identified by its table name, and optional instance identifier, and can be furthered filtered by its instance, columns, and row values. Additionally, a \texttt{TableReference} string can be recursive, so that each part of the string can be replaced with another \texttt{TableReference} that points to a different instance value.

\begin{table}[htbp]
  \centering
  \small
  \caption{Example TableReference strings.}
  \label{tab:patterns}
  \begin{tabularx}{\columnwidth}{%
      >{\raggedright\arraybackslash}p{6.4cm}  
      X                                       
    }
    \toprule
    \rowcolor{gray!30}
    \textbf{Example} & \textbf{Pattern} \\
    \midrule
    \textbf{Basic Examples} &  \\
    \midrule
    \texttt{<<config>>}                           & DataFrame        \\ \hline
    \texttt{<<config.batch\_size>>}        & Column    \\ \hline
    \texttt{<<config[function::<<function.name[index::0]>>]>>}      & Nested   \\ 
    \midrule
    \textbf{Common Data Operation Patterns} \\
    \midrule
    \texttt{<<users>>}                           & Reduction        \\ \hline
    \texttt{<<users[index::self.index]>>}        & One-to-One    \\ \hline
    \texttt{<<users[index::0:self.index]>>}                 & Cumulation   \\ \hline
    \texttt{<<users[index::self.index-5:self.index]>>}    & Convolution   \\ \hline
    \texttt{<<users[name::``alice'']>>}               & Selection     \\
    \bottomrule
  \end{tabularx}
\end{table}

Table \ref{tab:patterns} shows valid examples of the TableReference format. In functions with single-row output, a TableReference string supports a relative index, \texttt{``self.index''}, that references to the current row index being processed. More broadly, the \texttt{``self''} keyword is a general reference to the current instance being executed. As shown in the table, the relative index is useful because it can directly capture many different execution patterns common in data science (reduction, one-to-one, cumulation, convolution, selection)\cite{dslog}. There are additional metadata keywords that allow the operation  to access metadata within an instance.

These strings implicitly track the data lineage between different table instances. This is an initial design for TableVault to capture lineage universally for both execution and operation input configurations. In the future, we also aim to automatically support dynamic declaration so that data imports and lineage can be captured during the operation.

Since all executed code files, YAML builder files, and extracted data lineage can be retrieved by a \texttt{TableReference} string, any recorded information from previous data executions can be used in subsequent operations. We envision that this enables more informed dynamic workflows that can be self-organizing.

Now, the second way data can be ingested into the TableVault system is through direct imports. This is less ideal for full metadata tracking, but is required to support ingesting external datasets and external data modification tools.In this scenario where external tools or data are enabled, the worst case scenario is when data ingestion occurs, but is not recorded. For example, if an author might update a dataset with a version generated on a newer prompt, but forget to include that information in the metadata, it invalidates subsequent applications of the data. TableVault prevents this scenario by controlling the workflow's storage layer. All data ingestion into the repository must occur through its API, which enforces records of the operation to saved metadata. The existence of any external modifications becomes linked to a particular data artifact and author, mitigating some potential issues.

\subsection{Summary}

In general, TableVault is built on the principle that \textit{the existence} of all data artifacts and operations executed in collaborative human-model data workflows should be recorded, and the records themselves should be directly accessible as data artifacts. It is designed to encourage operations to be executed and data to be stored within its system for maximum traceability, while being flexible in supporting external data manipulations and data artifacts. This is not trivia, requiring insight into both the data storage, and operation execution. However, we envision that, as workflows are becoming more complex, there is an increasing need to record and organize metadata surrounding these workflows, and such organization can enable better human-model collaboration.

We leave additional details about the data execution and ingestion processes to the TableVault documentation\footnote{https://tablevault.org/}.

Outside of the current project scope, future challenges exist in considering the performance and scalability of the system, as well as considering the integration of different model interactions, e.g. potentially malicious models or more autonomous agents. Additionally, as the field matures, we envision that more standardized formats and practices may further impact how we consider metadata in these workflows \cite{MCP2025}.

\section{Workflow Case Study: Using TableVault to Categorize Documents}

To demonstrate intuition for the core TableVault API, we show a basic example of how the system could be used to call the OpenAI API to categorize a collection of documents. This type of data classification with AI models has been widely adopted and researched in many domains\cite{Zhang2023ClusterLLMLL, Cai2022SemanticenhancedIC, Liu2024OrganizingUI, Viswanathan2023LargeLM}, and is a good example of how various data artifacts may be stored and interacted with within a TableVault repository.

\subsection{Initial Setup}
An initial TableVault repository can be simply created by specifying a folder directory on disk:

\begin{lstlisting}
tv = TableVault("example_repository", author = "alice")
\end{lstlisting}

In the Python API, the \texttt{TableVault} object reuses the same repository. Each operation is linked to an author identifier; in the same workflow, this author can be a human actor, an AI model instance, or a parent operation process. 

For this task, we would like to create two tables with instances: \texttt{document-store} to store imported documents and \texttt{openai-response} to store OpenAI API responses.

\begin{lstlisting}
for name in ["document-store", "openai-response"]:
    tv.create_table(name)
    tv.create_instance(name)
\end{lstlisting}

Instances are not referenceable by a \texttt{TableReference} string until their internal dataframe is created. Each version of an instance is linked to a creation timestamp, and no further modifications are made after creation. In our use cases, we found that model execution was inherently the most expensive part of the workflow, and a more generous performance and storage budget would be reasonable if it allowed us to avoid recomputation. Backtracking between different versions was frequent to compare the effect of different input parameters, such as prompts. 

\subsection{Data Collection}

In order to import documents into the instance, we can specify a new builder file:

\begin{lstlisting}
tv.create_builder_file(table_name = "document-store", 
    builder_name = "document-store-index")
---
# user specified Builder file
builder_type: 'IndexBuilder'
changed_columns: ['file_name', 'artifact_name']
primary_key: ['file_name']
python_function: 'create_paper_table_from_folder'
code_module: 'table_generation'
is_custom: false
arguments:
    folder_dir: '../example_stories'
    artifact_folder: '~artifact_folder~'
dtypes:
  artifact_name: 'artifact_string'
\end{lstlisting}

The builder states that PDF files are imported from a local directory using the pre-existing \texttt{create\_paper\_table\_from\_folder} function from the \texttt{table\_generation} module. The \texttt{\textasciitilde artifact\_ folder\textasciitilde} string is a key word that informs the function where to store the imported files and is dynamically assigned by TableVault before execution. The system is responsible for the organization of files, but the parsing of the files is determined by the user program. This differs from some document processing systems that control data ingestion \cite{shankar2025docetlagenticqueryrewriting}, and allows more flexibility. 

We can execute the builder file with the following command:

\begin{lstlisting}
    tv.execute_instance("document-store")
\end{lstlisting}

\subsection{OpenAI API Calls}
Suppose we now create a code function that uploads a document to the OpenAI cloud storage and asks a model to classify the document. The pseudocode for such a function may look like:

\begin{lstlisting}
    tv.create_code_module("openai_helper")
---
# ask_openai function in document-store
def ask_openai(document, question, model):
    # upload document to openai cloud server
    file_id = openai.upload(document)
    # ask question
    response = openai.ask(model, question, file_id)
    return response
\end{lstlisting}

We would like to run this function on artifacts stored in the \texttt{documents} table and store the results in the \texttt{openai\_responses} table. This can be specified using the following two builder files:

\begin{lstlisting}
# user specified IndexBuilder 
builder_type: 'IndexBuilder'
changed_columns: ['file_name']
primary_key: ['file_name']
python_function: 'create_data_table_from_table'
code_module: 'table_generation'
is_custom: false
arguments:
    folder_dir: '<<document-store.file_name>>'
---
# User specified ColumnBuilder
builder_type: 'ColumnBuilder'
changed_columns: ['model_response']
python_function: 'ask_openai'
code_module: 'openai_helper'
nthreads: '<<parameters.threads[operation::~id~]>>'
arguments:
    document: '<<document_store.artifact_name[ file_name::<<self.file_name[self.index]>> ]>>'
    question: 'Is the story fiction or non-fiction? Reply with only one word.'
    model: '4o-mini'
\end{lstlisting}

The first builder file states that this table will have the same row index as the latest \texttt{document-store} table instance. The second builder file states that we would like to call the custom OpenAI function once for each row and fetch an imported artifact from \texttt{document-store} as an input argument. Since these operations are executed within TableVault, they are robustly tracked and can be safely paused, restarted, or canceled without unexpected side effects on the data within the repository. The operation status is stored at the same location as the data provenance, column-level metadata, and optional text descriptions. 

The configuration parameter \texttt{nthreads} is stored in another table in the repository that could be generated with TableVault by us, or other authors. Lineage for \texttt{nthreads} and \texttt{question} is both traced \texttt{TableReference} strings, and differentiation between execution optimization and operation input is up to the end user.

\subsection{Manual Human Edits}

After executing the previous table instance, we may see the resulting rows in the output dataframe:

\begin{center}
\small  
\renewcommand{\arraystretch}{1.1}  
\begin{tabular}{@{}r l l@{}}  
\toprule
 & \textbf{file\_name} & \textbf{model\_response} \\
\midrule
0 & \texttt{little\_red\_riding\_hood.pdf} & \texttt{fiction} \\
1 & \texttt{titanic.pdf}                  & \texttt{this is nonfiction} \\
\bottomrule
\end{tabular}
\end{center}

We can observe that the format was not respected in the second response. In this scenario, model accuracy is not relevant and we would like a standardized format for downstream processing. Therefore, we, an human author, may choose to manually inspect the dataframe, and directly modify the responses. A quick fix can be made directly using the Pandas library\cite{reback2020pandas}:

\begin{lstlisting}
df = tv.get_dataframe("openai-response")
df.loc[df['file_name'] == 'titanic.pdf', 'model_response'] = 'nonfiction'

tv.create_instance("openai-response", external=True)
tv.write_instance(df, "openai-response", description = "manual format corrections")
\end{lstlisting}

Since the exact execution is not seen by the TableVault library, the description text provides further context. The author should include the appropriate information and can be identified for further queries.

\subsection{Sketch: Automatic Generation of Tables}
In this example, documents were only categorized along a single dimension. However, suppose that we would like to implement a more complex workflow, where there exists a hierarchical tree of categories, and we would like to generate a new table that contains the classification of relevant document subsets for each vertex. Although each table could be created manually, this is inefficient and the direct lineage between the tables and the categories is lost. Instead, we could execute this workflow with a table instance that \textit{calls the TableVault API} within its operation, dynamically creating new tables and executing new instances.

The author of each generated table is set to the current instance's unique process identifier, so that execution origin is directly preserved. The parent instance operation can directly monitor the process of each child instance using its own TableVault instance, and make dynamic decisions about subsequent workflows. Each child would also have access to execution and lineage metadata about its parent for internal decisions.

In our current implementation, this process is partially supported, but requires the user to proactively provide more detailed metadata about operator decisions. In the future, we would like to make this process more automatic; we envision this dynamic recursive generation pattern as the basis for how more autonomous agents may interact with TableVault.
\section{Execution Cost}
The performance of workflows in TableVault has not yet been optimized, and we only present preliminary latency of singular operations to demonstrate that the system does not introduce any unexpected overhead. Each experiment was replicated three times, with the execution time of all trials shown.

\begin{figure}[htbp]
  \centering
  \includegraphics[width=\linewidth]{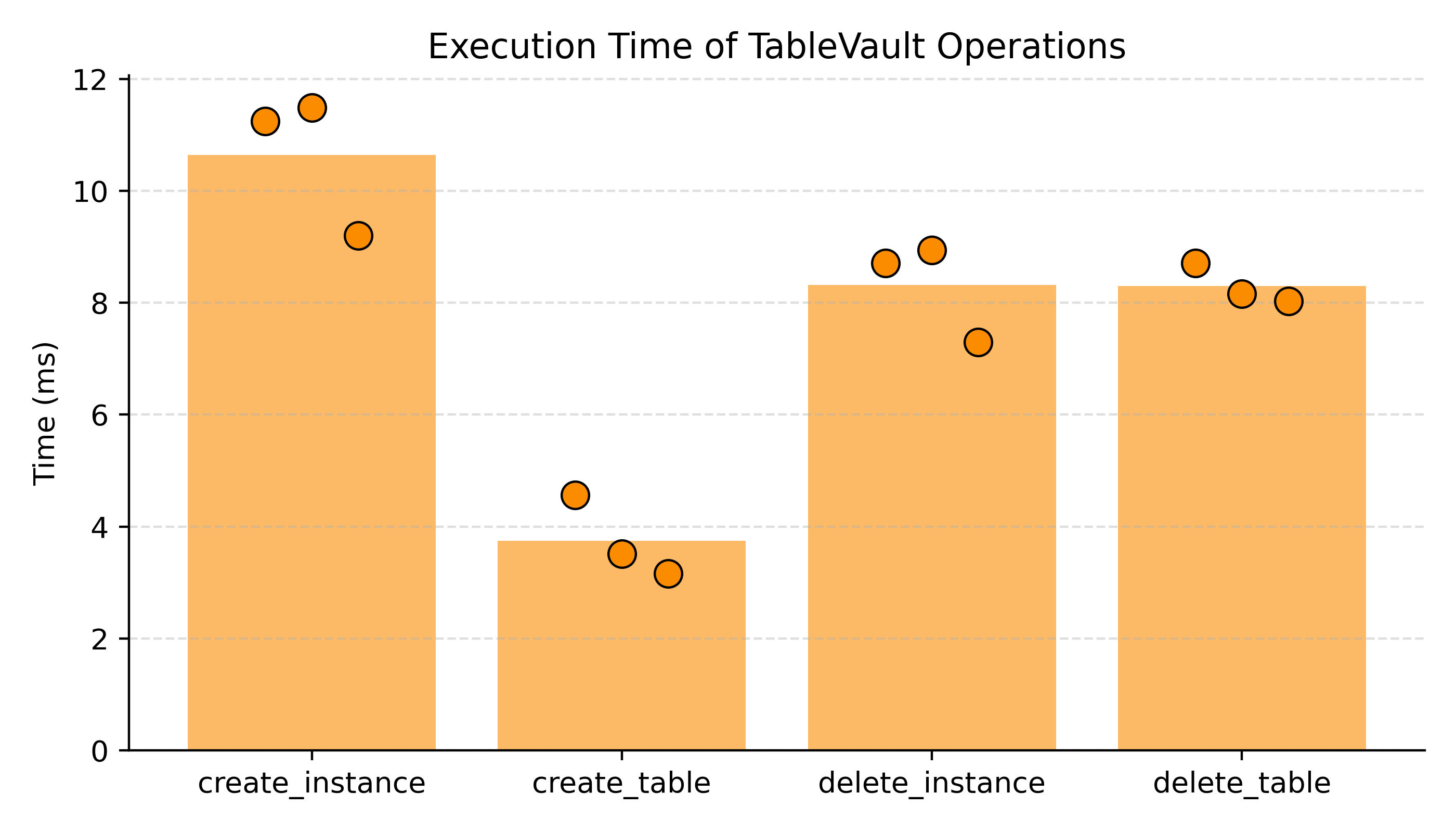}
  \caption{Execution time of basic TableVault operations.}
  \label{exp:functions}
\end{figure}

\begin{figure}[htbp]
  \centering
  \includegraphics[width=\linewidth]{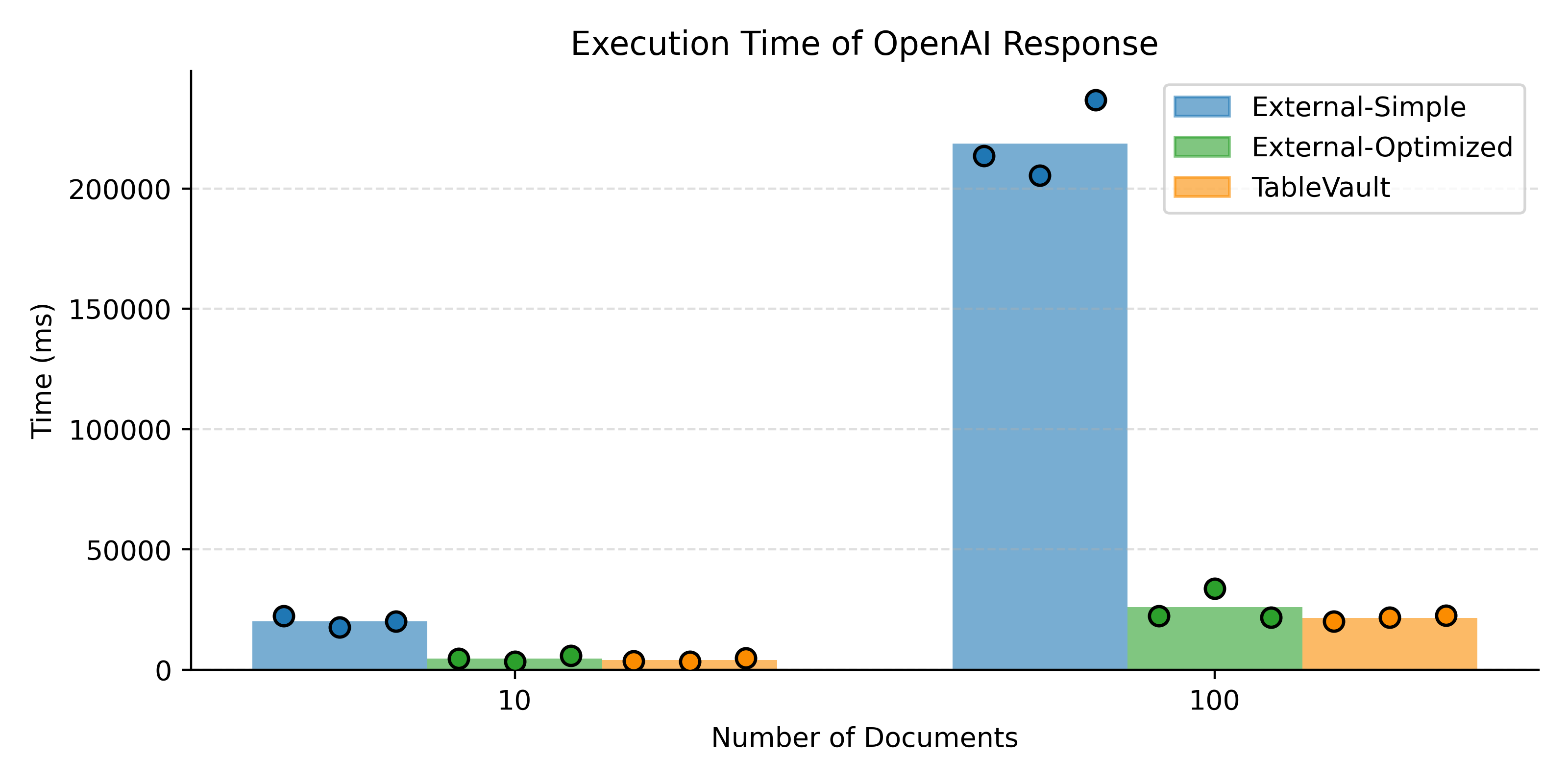}
  \caption{Execution time of OpenAI response over documents.}
  \label{exp:open-ai}
\end{figure}

Figure \ref{exp:functions} shows the execution time of various data modification operations in TableVault. We can observe that the cost of execution is on the order of milliseconds.

Figure \ref{exp:open-ai} shows the execution time of the OpenAI operation in our example compared to the pure Python baselines. The samples were generated by randomly selecting documents from the Grimm's Fairy Tales and Reuters-50-50 News datasets \cite{reuter_50_50_217, schomacker_grimms_2018}. For single-row operations, TableVault natively supports the option for multithreading, as well as incremental dataframe writes, with map-reduce data compilation. \textbf{External-Optimized} shows the performance of a Python program that was written to include those optimizations, and \textbf{External-Simple} shows the performance of a program that does not.

Both External-Optimized and TableVault performed significantly better than External-Simple. Although TableVault theoretically adds a slight overhead, the variation in API response time obscured any obvious delay.  
\section{Related Works}
Many recent systems have explored optimizing performance on LLM-integrated workflows. Lotus introduces semantic operators and applies a model cascade to those operations \cite{patel2025semanticoperatorsdeclarativemodel}. Palimspzet optimizes the performance of a logical plan while meeting accuracy and quality constraints \cite{liu2024declarativeoptimizingaiworkloads}. DocETL recursively applies operational optimization to a language model query while evaluating sampled data \cite{shankar2025docetlagenticqueryrewriting}. Aero adaptively manages predicate order and GPU utilization in queries that treat machine learning operations as a user0defined function \cite{Kakkar2025AeroAQ}. Abacus improves the original Palimspzet system through employing a multi-armed bandit algorithm to meet user specifications \cite{russo2025abacuscostbasedoptimizersemantic}.

In addition, frameworks have been proposed to standardize the way that language models interact with external tools and code\cite{Madden2024DatabasesUQ}. The core difference between these frameworks and TableVault is that TableVault focuses on transparency and leaves the physical execution to the user. DSPy integrates language model prompts within Python while largely abstracting the execution layer \cite{khattab2023dspycompilingdeclarativelanguage}. LangGraph enables agents by creating a directed graph of linked LLM operators \cite{langgraph_0_6_3}. CrewAI enables dynamic pipeline orchestration by allowing agents to spawn subprocesses \cite{crewai_2025}. Aryn models documents as hierarchical entities and exposes a logical plan for debugging purposes \cite{anderson2024designllmpoweredunstructuredanalytics}.  

Another set of relevant systems explores how metadata and provenance can be captured and stored in data systems. Ground proposes a centralized service that consolidates metadata from many different applications \cite{Hellerstein2017GroundAD}. MLInspect annotates data operations at runtime to evaluate data distribution shifts \cite{Grafberger2021LightweightIO}. CrowdER extends SQL to accommodate crowdsourcing information within a database schema \cite{Franklin2011CrowdDBAQ}. Smoke captures fine-grained, row-level provenance over seven classical SQL operations with lightweight lineage instrumentation \cite{Psallidas2018SmokeFL}.

\section{Conclusion}

In this paper, we present the initial design of the TableVault framework for metadata capture in AI-augmented data workflows.

As AI models develop, there are exciting questions about how they should influence not only how metadata is tracked but also how metadata should be ultimately consumed. Fundamentally, these models enable more complexity in data workflow patterns, and we envision future work beyond this paper in establishing how metadata can be used to support this complexity, while promoting good data practices.  

\bibliographystyle{ACM-Reference-Format}
\bibliography{references}

\end{document}